\documentclass[9pt,twocolumn,twoside]{osajnl}
\journal{ol} % Choose journal (ao, aop, josaa, josab, ol, optica, pr)

\setboolean{shortarticle}{true}
\title{Thermally Induced Generation of Platicons in Optical Microresonators}

\author[1,*]{Valery~E.~ Lobanov}
\author[1]{Nikita~M.~ Kondratiev}
\author[1,2]{Igor~A.~Bilenko}

\affil[1]{Russian Quantum Center, 143026, Skolkovo, Russia}
\affil[2]{Faculty of Physics, Lomonosov Moscow State University, 119991, Moscow, Russia}

\affil[*]{Corresponding author: v.lobanov@rqc.ru}
\begin{abstract}
We demonstrate numerically novel mechanism providing generation of the flat-top solitonic pulses, platicons, in optical microresonators at normal GVD via negative thermal effects. We found that platicon excitation is possible if the ratio of the photon lifetime to the thermal relaxation time is large enough. We show that there are two regimes of the platicon generation depending on the pump amplitude: the smooth one and the oscillatory one. Parameter ranges providing platicon excitation are found and analysed for different values of the thermal relaxation time, frequency-scan rate and GVD coefficient. Possibility of the turn-key generation regime is also shown.

\end{abstract}

\setboolean{displaycopyright}{true}

\begin{document}

\maketitle

A manifestation of thermal effects, such as thermo-optic and thermal expansion effects, is inevitable in modern microresonator platforms \cite{Ilchenko1992ThermalNE,Fomin:05}. Microresonator thermal effects are often considered as parasitic, especially in the context of optical nonlinear processes, where thermally induced drifts, fluctuations and instabilities \cite{Fomin:05,Carmon:04,Grudinin:09,Diallo:15,PhysRevA.103.013512} can strongly impact the generation of optical frequency combs and solitons \cite{herr2014temporal,Bao:17}. Different methods were developed for the compensation of these effects including precise adjustment of the frequency-scan rate \cite{herr2014temporal,PhysRevLett.121.063902}, various schemes of the pump frequency and pump power modulation \cite{Wildi:19,Li:17,Brasch357,Brasch:16}, active feedback loops \cite{Yi:16}, application of auxiliary laser \cite{Grudinin:11,Zhang:19,Zhou2019} or even cryogenic temperatures \cite{PhysRevApplied.12.034057}. On the other hand, these effects were also sometimes beneficial, allowing a generation of solitons without frequency scan \cite{Tanabe2016} 
and stable access to the single-soliton regime \cite{Guo2017}. Using additional heater one may also control generation of the dissipative Kerr solitons \cite{Joshi:16} and tune comb parameters \cite{Xue:16}. While the impact of thermal effects on the generation of optical frequency combs was investigated from a number of perspectives, the majority of the works has been focused on microresonator systems with anomalous group velocity dispersion (GVD) generating bright solitons. Despite the growing interest to the generation of optical frequency combs in cw-laser-driven normal-GVD microresonators, the influence of thermal effects on these systems have not been studied in details to date. 

Here, we numerically investigate the impact of thermal effects on nonlinear optical processes in microresonators with normal GVD. We first demonstrate that thermal effects can solely induce the generation of the flat-top solitonic pulses, platicons \cite{Lobanov2015} (or dark solitons \cite{PhysRevA.89.063814}). This is usually a challenging task requiring specific mode structure \cite{Lobanov2015,Xue2015,Jang:16}, complex multi-resonator settings \cite{Kim:19}, pump modulation \cite{Lobanov2015epl,Lobanov2019} or self-injection locking regime \cite{KondratievNum:20,jin2020hertzlinewidth}. Generation of platicons was shown to be more efficient in terms of the pump-to-comb conversion efficiency than the generation of bright solitons \cite{Kim:19}. We found that while positive thermal effects (thermal resonance shift has the same direction as nonlinear shift) lead to the resonance shift only, at negative thermal effects platicon excitation may be realized upon the pump frequency scan without any additional approaches if the ratio of the thermal relaxation time and photon lifetime is small enough. It means that microresonator quality factor should be large enough to satisfy this condition. 
%Calcium fluoride microresonators possessing high quality factor (up to $10^{11}$ \cite{Savchenkov:07}) and negative thermo-optic coefficient \cite{Tanabe2016} seem to be a potential candidate for the observation of this effect.
Parameter range providing platicon generation was found and two distinct regimes of platicon generations were identified – the smooth one, and the oscillatory one. Moreover, we revealed the possibility of the turn-key regime of platicon generation enabled by the negative thermal effects.

For numerical simulations we used the system of the Lugiato-Lefever equation \cite{PhysRevA.89.063814} for the slowly varying envelope of the intracavity field $\Psi$ with the rate equation for the normalized thermally induced detuning $\Theta$ \cite{Carmon:04,PhysRevA.103.013512,herr2014temporal,Diallo:15,Grudinin:09,Grudinin:11}:
\begin{gather}
    \label{ThermalLLE}
    \begin{cases}
        \frac{\partial\Psi}{\partial\tau}=\frac{d_2}{2}\frac{\partial^2\Psi}{\partial\varphi^2}-[1+i(\alpha-\Theta)]\Psi+i|\Psi|^2\Psi+f,\\
        \frac{\partial\Theta}{\partial\tau}=\frac{2}{\kappa t_T}\left(\frac{n_{2T}}{n_2}\frac{U}{2\pi}-\Theta\right). \\ 
    \end{cases}
\end{gather}
Here $\tau=\kappa t/2$ denotes the normalized time, 
$\kappa=\omega_0/Q$ is the cavity total decay rate ($Q$ is the loaded quality factor), 
$\omega_0$ is the pumped mode resonant frequency, 
$\varphi\in[-\pi;\pi]$ is an azimuthal angle in the coordinate frame rotating with the rate equal to the microresonator free spectral range (FSR) $D_1$, 
$d_2=2D_2/\kappa$ is the normalized GVD coefficient, positive for the anomalous GVD and negative for the normal GVD [the microresonator eigenfrequencies are assumed to be $\omega_\mu = \omega_0+D_1\mu+\frac{D_2}{2}\mu^2$ %\textcolor{red}{[high-order dispersion terms are neglected]}
, where $\mu$ is the mode number, calculated from the pumped mode], 
$\alpha=2(\omega_0-\omega_p)/\kappa$ is the normalized detuning from the pump frequency $\omega_p$ from the pumped resonance.
The normalized pump amplitude for matched beam area and coupler refraction is $f=\sqrt{\frac{8\omega_{\rm p} c n_2\eta P_{\rm in}}{\kappa^2n^2 V_{\rm eff}}}$ \cite{herr2014temporal,Kondratiev:20}, where $c$ is the speed of light, 
$n_2$ is the microresonator nonlinear index, 
$P_{\rm in}$ is the input pump power, 
$n$ is the refractive index of the microresonator mode, 
$V_{\rm eff}$ is the effective mode volume, 
$\eta$ is the coupling efficiency [$\eta=1/2$ for critical coupling, $\eta\rightarrow1$ for overloaded], $U=\int|\Psi|^2d\varphi$. %Note that $f=1$ corresponds to the threshold power of the parametric four-wave mixing.

The normalized thermally induced resonance shift comes from thermo-optic  $\Theta=\frac{2}{n}\frac{dn}{dT}Q\delta T$ or thermal expansion effect $\Theta=2 \alpha_L Q\delta T$, where $\delta T$ is the temperature variation, $\alpha_L$ is thermal expansion coefficient, $t_T$ is the thermal relaxation time, $n_{2T}$ is the effective coefficient of the thermal nonlinearity \cite{Fomin:05,Ilchenko1992ThermalNE,herr2014temporal}. The ratio $n_{2T}/n_2$ defines the value of the thermally induced resonance shift for the considered intracavity power $U$; the sign of the thermal nonlinearity is defined by the sign of the thermo-optic coefficient $\frac{1}{n}\frac{dn}{dT}$ or thermal expansion coefficient $\alpha_L$. The thermal parameters $t_T$ and $n_{2T}$ both depend on material properties and the geometry of the resonator and are generally different for thermal refraction and thermal expansion \cite{Ilchenko1992ThermalNE,Grudinin:09,Diallo:15}. However, one of the effects can be negligible or significant depending on the setup \cite{Tanabe2016,Savchenkov_2018,Lim2017}. So, for the first estimations we consider a single composite effect, following \cite{Carmon:04,Grudinin:09,PhysRevA.103.013512} for simplicity. Interestingly, it is evident from the second equation of \eqref{ThermalLLE} that the absolute value of the thermal relaxation time is not so important as its relation to the photon lifetime $t_{\rm ph}$, since $2 / \kappa t_T=2t_{\rm ph} / t_T$. %\textcolor{red}{It should be noted that, e.g. similar to \cite{PhysRevA.103.013512}, we did not consider the combination of different thermal effects, naturally having different values of $t_T$ and $n_{2T}$, in our model.}

\begin{figure}[t!]
\centering\includegraphics[width=1\linewidth]{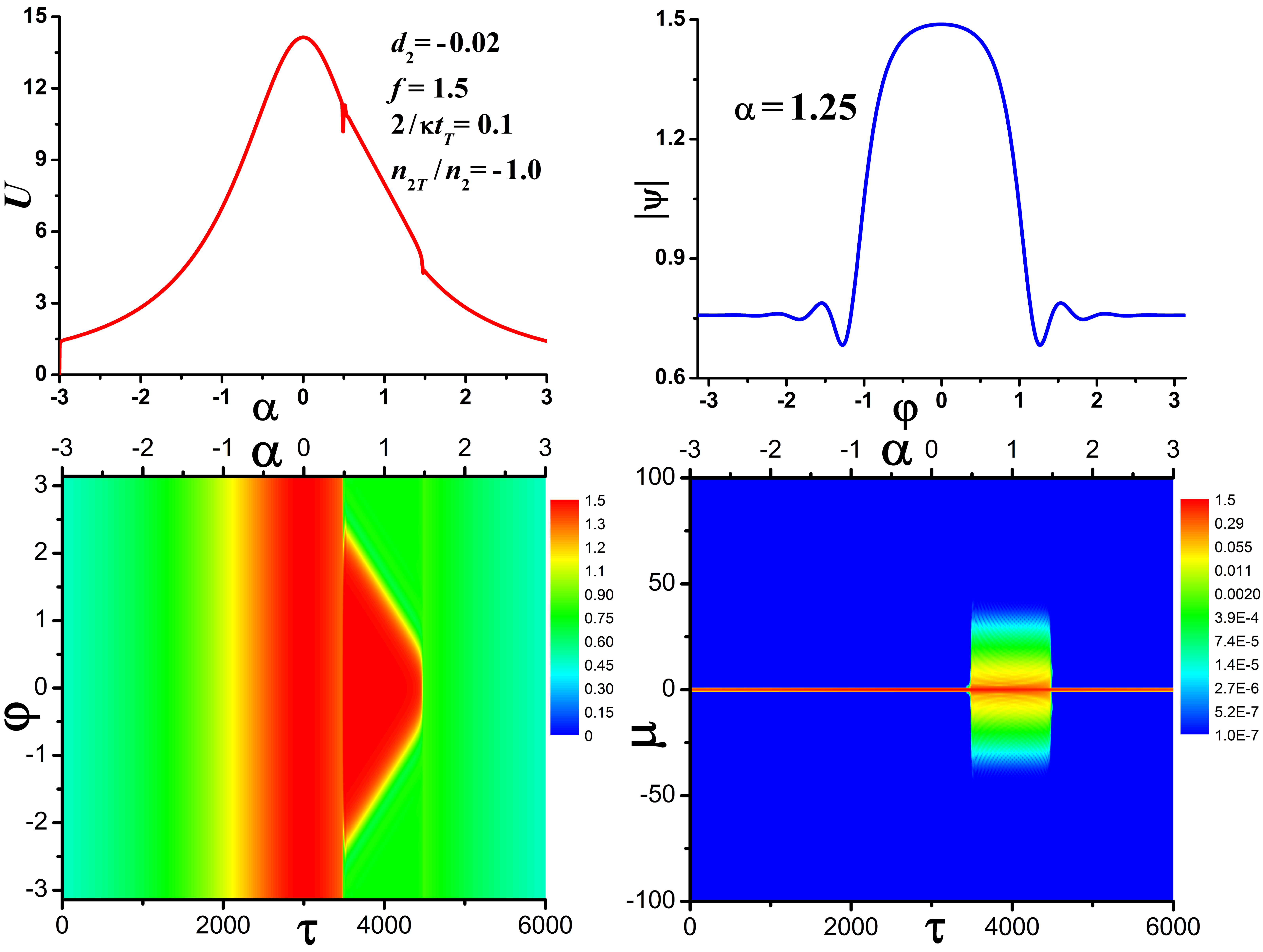}
\caption{Smooth generation regime at $2/\kappa t_T=0.1$, $n_{2T}/n_2=-1.0$, $d_2=-0.02$, $f=1.5$ and $v=0.001$. Top left: intracavity power $U$ vs pump frequency detuning $\alpha$ upon pump frequency scan. Top right: platicon profile at $\alpha=1.4$. Bottom panels: (left) field modulus and (right) spectrum evolution.
%Bottom left: field distribution evolution. Bottom right: spectrum evolution. 
\label{fig1}}
\end{figure}

We considered the case of the normal GVD ($d_2=-0.02$). First, we simulated nonlinear effects arising upon the linear-in-time pump frequency scan $\alpha(\tau)=\alpha(0)+vt$ for the different values of the pump amplitude $f$, normalized thermal relaxation rate $2/\kappa t_T$ and normalized thermal nonlinearity coefficient $n_{2T}/n_2$. At each step, we first calculated $\Theta$ for the known value of the intracavity power $U$ and then we solved the equation for $\Psi$ with obtained $\Theta$ using the standard split-step Fourier routine. 

\begin{figure}[t!]
\centering\includegraphics[width=1\linewidth]{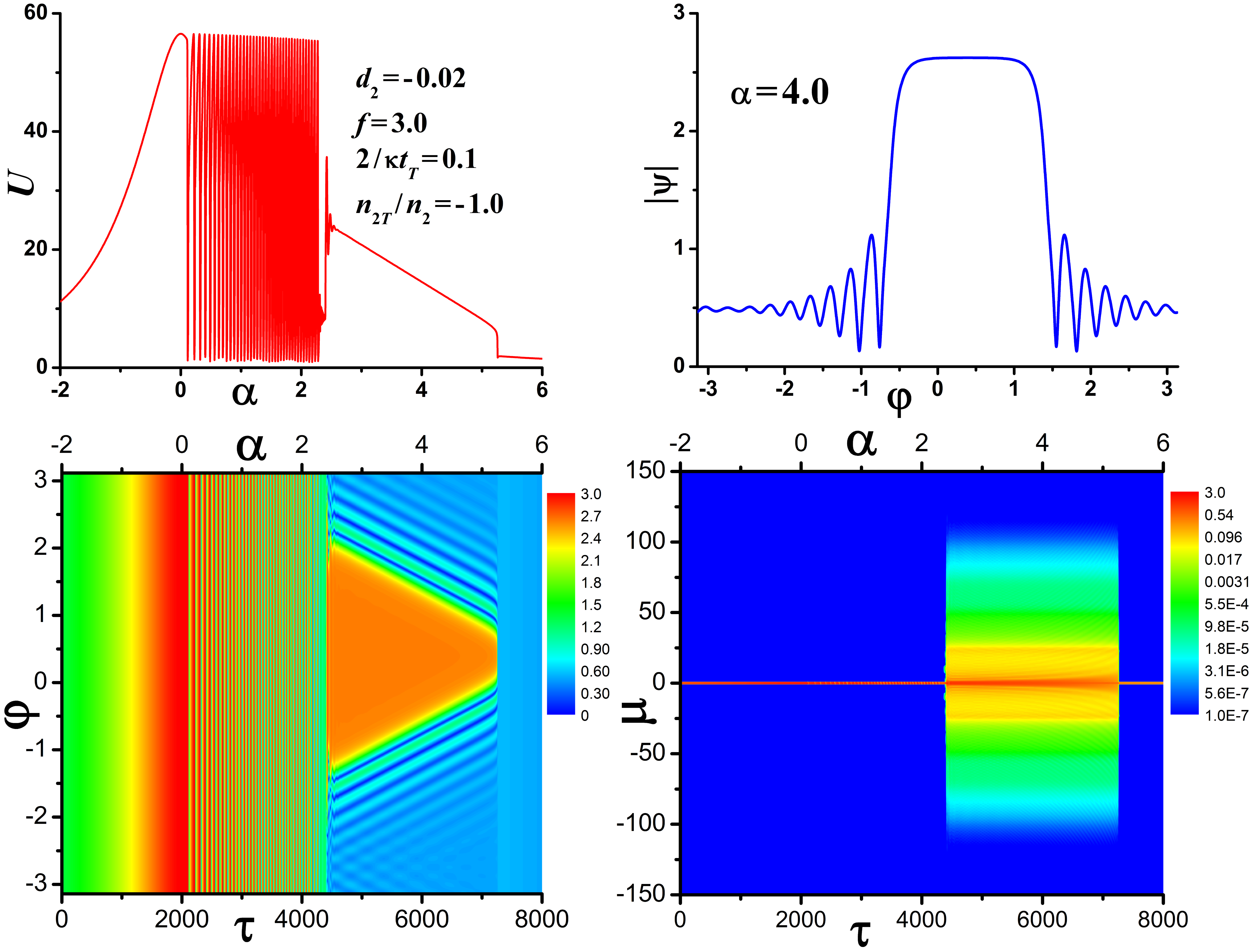}
\caption{Oscillatory generation regime at $2/\kappa t_T=0.1$, $n_{2T}/n_2=-1.0$, $d_2=-0.02$, $f=3.0$ and $v=0.001$. Top left: $U$ vs $\alpha$ upon pump frequency scan. Top right: platicon profile at $\alpha=4.0$. Bottom panels: (left) field modulus and (right) spectrum evolution. %Bottom left: field distribution evolution. Bottom right: spectrum evolution.
\label{fig2}}
\end{figure}

For the positive thermal effect $n_{2T}/n_2>0$ we observed conventional triangle resonances without frequency comb generation. However, if the thermal effects were negative, the scan velocity and the thermal relaxation to photon lifetime ratio were small enough, the generation of platicons was possible in the specific range of the parameters. We revealed two regimes of platicon generation. The first one - smooth generation regime - is a transition from a cw solution to a platicon at small pump amplitudes $f$ (Fig. \ref{fig1}) that takes place at small values of $|n_{2T}/n_2|$ for a narrow range of $f$ depending on the particular thermal nonlinearity value. For $v=0.001$ and $2/\kappa t_T=0.1$ it was observed at $|n_{2T}/n_2|\in[0.5;2.1]$ and $f\in[1.3;2.1]$. One can see small variation of the resonance curve inclination in the top left panel in Fig. \ref{fig1} corresponding to the platicon generation. This modification range coincides with the range of the abrupt transformation of the field distribution [bottom left panel in Fig. \ref{fig1}] and spectrum [bottom right panel in Fig. \ref{fig1}]. We also checked that when frequency scan stops generated platicon propagates in a stable manner. 
\begin{figure}[h!]
\centering\includegraphics[width=1.0\linewidth]{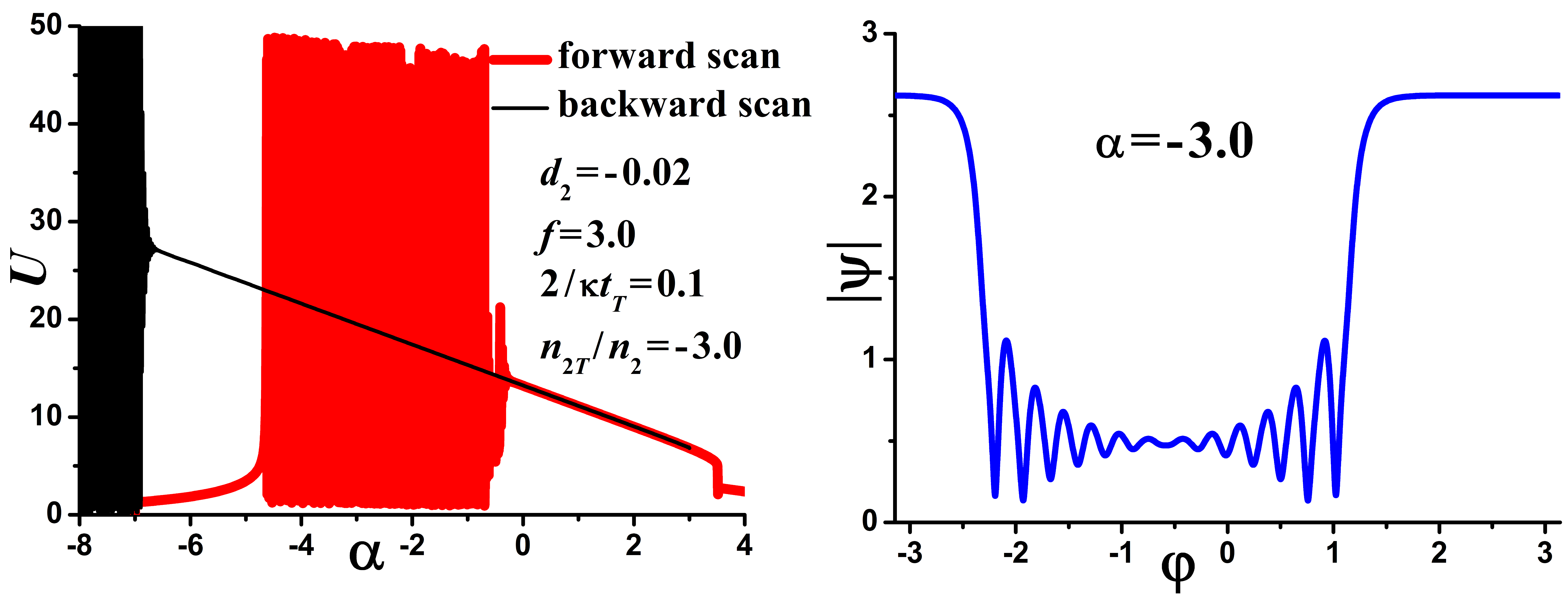}
\caption{Left: $U$ vs $\alpha$ upon forward pump frequency scan from the noise-like input [red line] and backward scan from the platicon input [black line] at  $2/\kappa t_T=0.1$, $n_{2T}/n_2=-3.0$, $d_2=-0.02$, $f=3.0$ and $v=0.001$. Right: profile of the platicon generated upon backward scan at $\alpha=-3.0$.\label{fig3}}
\end{figure}

\begin{figure*}[ht!]
\centering\includegraphics[width=0.9\linewidth]{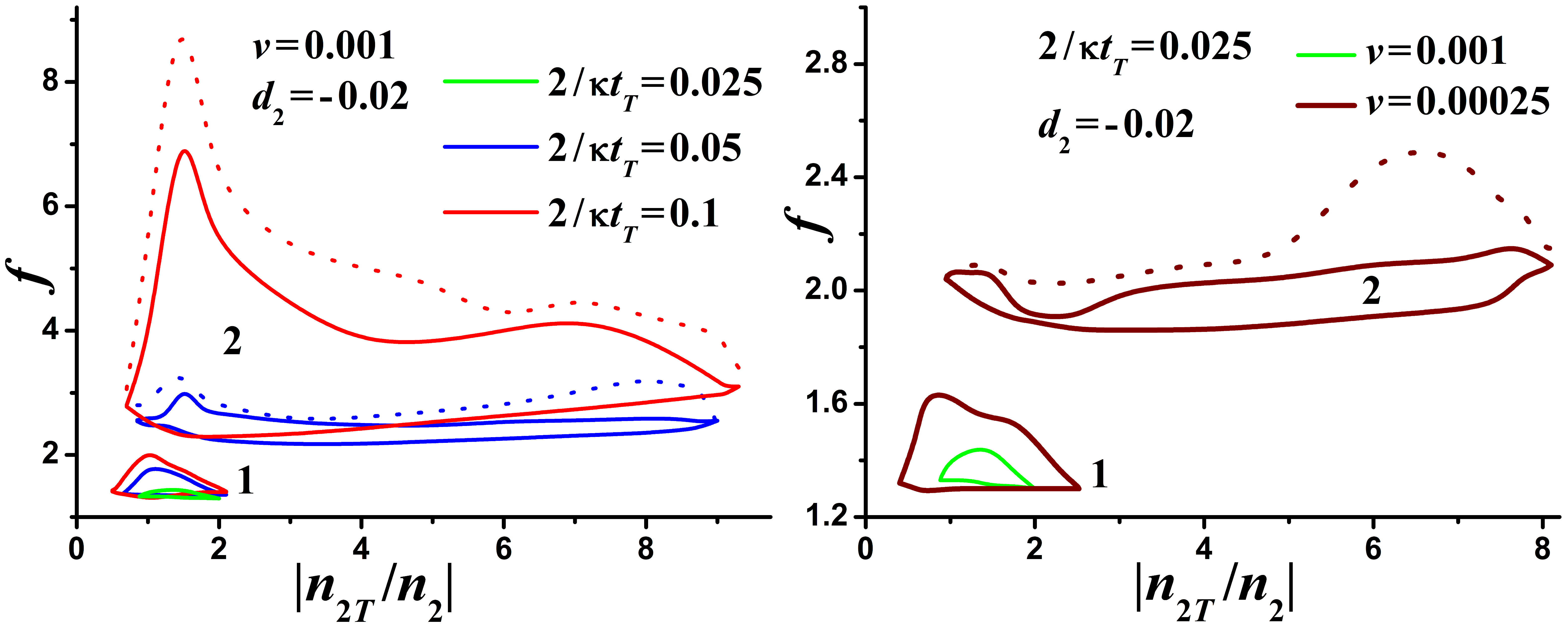}
\caption{Left panel: Generation domains for different values of $2/\kappa t_T$ at $d_2=-0.02$ and $v=0.001$. Domain 1 corresponds to the smooth generation regime, domain 2 – to the oscillatory generation regime. Different colors correspond to the different values of the thermal relaxation time. Between the solid and the dotted line of the same color platicon generation is possible but the result is unpredictable. Right panel: Generation domains for different scan rates $v$ at $d_2=-0.02$ and $2/\kappa t_T=0.025$.\label{fig4}}
\end{figure*}

Then, for larger values of the pump amplitude $f$ lying outside smooth generation range, we observed the transformation of the platicon generation into single-mode oscillations turning into cw solution upon further scan. Note, that these anharmonic oscillations of the intracavity power are regular and continue after the scan stops.

Surprisingly, even for higher values of the pump amplitude we found the second regime of platicon generation - oscillatory generation regime. Here the platicon emerges after an oscillatory transient process [see Fig. \ref{fig2}]. Note, that critical values of the pump amplitude for each regime depend on the combination of the scan rate and thermal relaxation time.

We also found that if scan is stopped in the platicon regime and reversed, the oscillations are either absent or begin at much lower detuning values and, thus, platicon can be generated at significantly wider range in terms of the detuning $\alpha$ than that observed upon forward scan only [see Fig. \ref{fig3}].

We defined the range of parameters providing platicon generation in coordinates that determine the thermal and nonlinear dynamics of the system, "normalized thermal nonlinearity coefficient $n_{2T}/n_2$ - dimensionless pump amplitude $f$", for different values of the normalized thermal relaxation time [see left panel in Fig. \ref{fig4} where different colors correspond to the different values of $2/\kappa t_T$]. Note, that for the oscillatory regime the growth of the pump amplitude first leads to the transformation of the deterministic platicon generation into probabilistic regime [critical values of the pump amplitude are shown by the solid lines in the domains marked with number 2 in Fig. \ref{fig4}] and then to the deterministic absence of the platicon generation [critical values are shown by the dotted lines of the same color in the same domains]. To determine these boundaries accurately we generated several realizations of the noise-like input for the different combinations of $n_{2T}/n_2$ and $f$ and simulated the processes upon the pump frequency scan. 

Also, it was found that generation domains become narrower with the growth of the thermal relaxation time [or with decrease of $2/\kappa t_T$]. This is more noticeable for the region corresponding to the oscillatory generation regime [marked with number 2 in Fig. \ref{fig4}]. For $2/\kappa t_T=0.025$ and $v=0.001$ [green lines in the left panel in Fig. \ref{fig4}] oscillatory regime was not observed. However, scan velocity decrease allows to compensate thermal time growth: at $2/\kappa t_T=0.025$ and $v=0.00025$ two platicon generation regimes were  observed again [right panel in Fig. \ref{fig4}].

\begin{figure}[h!]
\centering\includegraphics[width=0.9\linewidth]{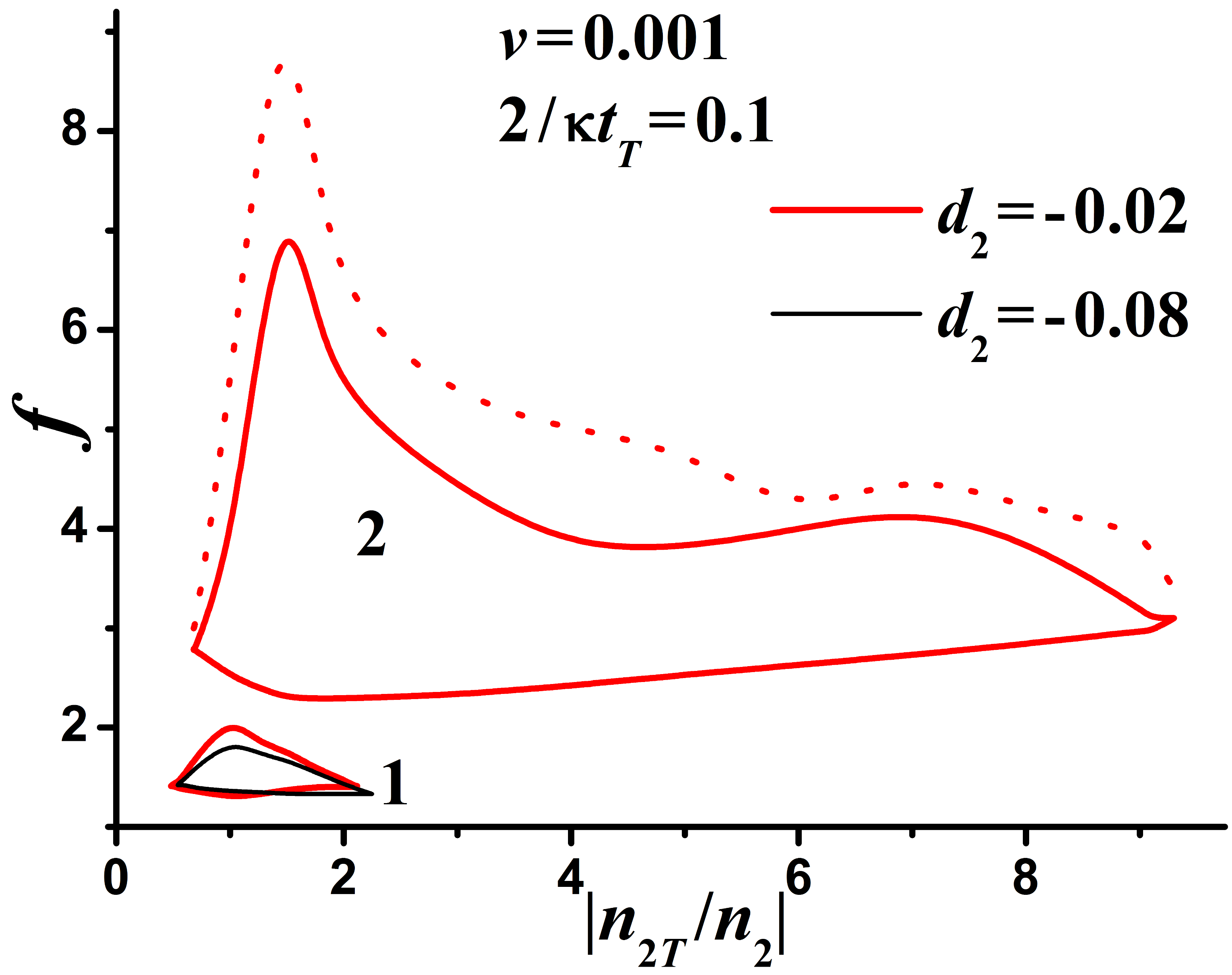}
\caption{Generation domains for different values of the GVD coefficient $d_2$ at $v=0.001$, $2/\kappa t_T$=0.1. Domain 1 corresponds to the smooth regime, domain 2 – to the oscillatory regime. \label{fig5new}}
\end{figure}

We also found that the generation domain becomes narrower with the growth of the dispersion coefficient modulus. For example, we observed widening of the both generation domains compared with those in Fig. \ref{fig4} if the GVD coefficient modulus was decreased from 0.02 to 0.01 and did not find the oscillatory regime if it was increased from 0.02 to 0.08 [see Fig. \ref{fig5new}]. So, to maintain the platicon generation, the increase of the GVD coefficient absolute value requires the decrease of the pump frequency scan velocity. Interestingly, at high values of the GVD coefficient and small scan rate [e.g. at $v=0.0001$ , $d_2=-0.08$, $2/\kappa t_T=0.1$, $n_{2T}/n_2=-4.0$] we observed transformation of the generated platicon into breathing platicon upon further frequency scan.

\begin{figure}[h!]
\centering
\includegraphics[width=1.0\linewidth]{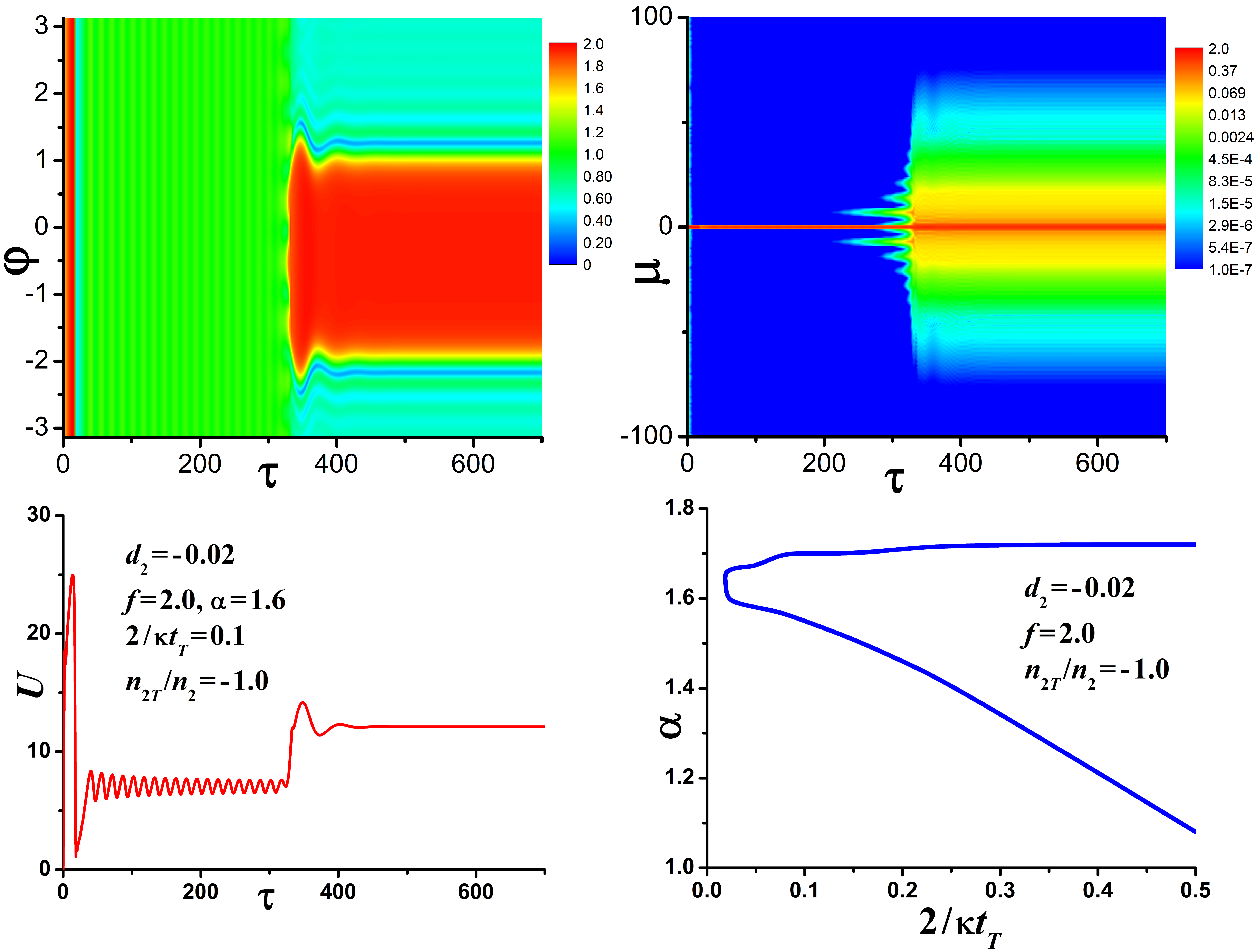}

\caption{Generation of platicons without frequency scan at $n_{2T}/n_2=-1.0$ $d_2=-0.02$ and $f=2.0$. Top panels: (left) field modulus and (right) spectrum evolution at $2/\kappa t_T=0.1$ and $\alpha=1.6$. Bottom left: $U$ vs $\tau$ upon propagation. Bottom right: range of pump frequency detuning $\alpha$ providing platicon generation without frequency scan vs $2/\kappa t_T$.\label{fig6}}
\end{figure}

Also, we revealed that in some cases platicons can be generated from the noise-like input without frequency scan [see Fig. \ref{fig6}] that opens the way to the turn-key microresonator comb system. This allows to use more stable fixed-frequency laser as a pump source and to simplify significantly comb generation system by not using complex tuning and control schemes. The detuning range providing platicon generation becomes narrower with the growth of the thermal relaxation time [see Fig. \ref{fig6}, bottom right].

%Calcium fluoride seems to be a perspective material for the realization of the considered process since it combines both negative thermo-optic coefficient \cite{Tanabe2016} and high Q-factor [up to $10^{11}$ at $\lambda=1.55$ $\mu$m \cite{Savchenkov:07}] providing large enough photon lifetime. In \cite{Tanabe2016} it was reported $t_T^{-1}=1.28\times10^{4} s^{-1}$, thus in order to obtain $2/\kappa t_T>0.025$ at $\lambda=1.55$ $\mu$m one needs $Q>2.5\times10^{9}$ that seems to be realistic. Also, calcium fluoride has zero-dispersion point at $\lambda=1.46$ $\mu$m, thus it is possible to obtain weak normal dispersion close to this wavelength via dispersion engineering \cite{FujiiTanabe2020}. 
According to our preliminary estimations based on the considered simplified model, revealed conditions for the platicon excitation via thermal effects seem to be experimentally feasible. In order to obtain $2/\kappa t_T>0.025$ at $\lambda=1.55$ $\mu$m for the thermal relaxation time of $100 \mu s$ one needs $Q>1.95\times10^{9}$ that is possible in crystalline microresonators and nearly possible in on-chip systems. According to our numerical results dimensionless scan velocity $v$ should be less than 0.001. One can easily calculate real pump frequency scanning speed from the normalized value $v$ using simple formula: $\frac{d\omega_p/2\pi}{dt}=v\frac{\pi c^2}{2\lambda_p^2Q^2}$. For $Q=5\times10^9$ and $\lambda_p=1.55$ $\mu$m and $v=0.001$ corresponds to the scan rate of 3.65 MHz/s. Small enough GVD coefficient  could be realized via dispersion engineering \cite{FujiiTanabe2020}.
However, a more accurate analysis with both thermal effects having different parameters \cite{Tanabe2016,Diallo:15} should be carried out for particular setup, especially for bulk crystalline microresonators, e.g. for $CaF_2$ ones, which have negative thermo-optic coefficients and high Q-factors [up to $10^{11}$ at $\lambda=1.55$ $\mu$m \cite{Savchenkov:07}] but also experience significant positive thermal expansion effect  \cite{Tanabe2016}. This is an important topic of further research.

To summarize, we demonstrated numerically novel mechanism of the platicon generation in optical microresonators at normal GVD via thermal effects. We found that platicon excitation is possible if thermal effect is negative and the ratio of the photon lifetime and thermal relaxation time is large enough. We showed that there are two regimes of platicon generation depending on the pump amplitude: smooth one and oscillatory one. We defined parameter ranges providing platicon excitation and analysed them for different scan rates and GVD values. The turn-key generation regime was also demonstrated.

\begin{backmatter}
\bmsection{Funding} Russian Science Foundation (Project No. 17-12-01413-$\Pi$).
\bmsection{Acknowledgments} %V.E.L. acknowledges personal support from the Foundation for the Advancement of Theoretical Physics and Mathematics “BASIS”.
V.E.L. acknowledges personal support from the Foundation for the Advancement of Theoretical Physics and Mathematics “BASIS”.
\bmsection{Disclosures} The authors declare no conflicts of interest.
\end{backmatter}

% Bibliography
\bibliography{sample}

\begin{thebibliography}{10}
\newcommand{\enquote}[1]{``#1''}

\bibitem{Ilchenko1992ThermalNE}
V.~Ilchenko and M.~L. Gorodetskii, \enquote{Thermal nonlinear effects in
  optical whispering gallery microresonators,} {\protect\JournalTitle{Laser
  Physics}} \textbf{2}, 1004--1009 (1992).

\bibitem{Fomin:05}
A.~E. Fomin, M.~L. Gorodetsky, I.~S. Grudinin, and V.~S. Ilchenko,
  \enquote{Nonstationary nonlinear effects in optical microspheres,}
  {\protect\JournalTitle{J. Opt. Soc. Am. B}} \textbf{22}, 459--465 (2005).

\bibitem{Carmon:04}
T.~Carmon, L.~Yang, and K.~J. Vahala, \enquote{Dynamical thermal behavior and
  thermal self-stability of microcavities,} {\protect\JournalTitle{Opt.
  Express}} \textbf{12}, 4742--4750 (2004).

\bibitem{Grudinin:09}
I.~S. Grudinin and K.~J. Vahala, \enquote{Thermal instability of a compound
  resonator,} {\protect\JournalTitle{Opt. Express}} \textbf{17}, 14088--14098
  (2009).

\bibitem{Diallo:15}
S.~Diallo, G.~Lin, and Y.~K. Chembo, \enquote{Giant thermo-optical relaxation
  oscillations in millimeter-size whispering gallery mode disk resonators,}
  {\protect\JournalTitle{Opt. Lett.}} \textbf{40}, 3834--3837 (2015).

\bibitem{PhysRevA.103.013512}
A.~Leshem, Z.~Qi, T.~F. Carruthers, C.~R. Menyuk, and O.~Gat, \enquote{Thermal
  instabilities, frequency-comb formation, and temporal oscillations in {K}err
  microresonators,} {\protect\JournalTitle{Phys. Rev. A}} \textbf{103}, 013512
  (2021).

\bibitem{herr2014temporal}
T.~Herr, V.~Brasch, J.~D. Jost, C.~Y. Wang, N.~M. Kondratiev, M.~L. Gorodetsky,
  and T.~J. Kippenberg, \enquote{Temporal solitons in optical microresonators,}
  {\protect\JournalTitle{Nat. Photon.}} \textbf{8}, 145--152 (2014).

\bibitem{Bao:17}
C.~Bao, Y.~Xuan, J.~A. Jaramillo-Villegas, D.~E. Leaird, M.~Qi, and A.~M.
  Weiner, \enquote{Direct soliton generation in microresonators,}
  {\protect\JournalTitle{Opt. Lett.}} \textbf{42}, 2519--2522 (2017).

\bibitem{PhysRevLett.121.063902}
J.~R. Stone, T.~C. Briles, T.~E. Drake, D.~T. Spencer, D.~R. Carlson, S.~A.
  Diddams, and S.~B. Papp, \enquote{Thermal and nonlinear dissipative-soliton
  dynamics in {K}err-microresonator frequency combs,}
  {\protect\JournalTitle{Phys. Rev. Lett.}} \textbf{121}, 063902 (2018).

\bibitem{Wildi:19}
T.~Wildi, V.~Brasch, J.~Liu, T.~J. Kippenberg, and T.~Herr, \enquote{Thermally
  stable access to microresonator solitons via slow pump modulation,}
  {\protect\JournalTitle{Opt. Lett.}} \textbf{44}, 4447--4450 (2019).

\bibitem{Li:17}
Q.~Li, T.~C. Briles, D.~A. Westly, T.~E. Drake, J.~R. Stone, B.~R. Ilic, S.~A.
  Diddams, S.~B. Papp, and K.~Srinivasan, \enquote{Stably accessing
  octave-spanning microresonator frequency combs in the soliton regime,}
  {\protect\JournalTitle{Optica}} \textbf{4}, 193--203 (2017).

\bibitem{Brasch357}
V.~Brasch, M.~Geiselmann, T.~Herr, G.~Lihachev, M.~H.~P. Pfeiffer, M.~L.
  Gorodetsky, and T.~J. Kippenberg, \enquote{Photonic chip{\textendash}based
  optical frequency comb using soliton {C}herenkov radiation,}
  {\protect\JournalTitle{Science}} \textbf{351}, 357--360 (2016).

\bibitem{Brasch:16}
V.~Brasch, M.~Geiselmann, M.~H.~P. Pfeiffer, and T.~J. Kippenberg,
  \enquote{Bringing short-lived dissipative {K}err soliton states in
  microresonators into a steady state,} {\protect\JournalTitle{Opt. Express}}
  \textbf{24}, 29312--29320 (2016).

\bibitem{Yi:16}
X.~Yi, Q.-F. Yang, K.~Y. Yang, and K.~Vahala, \enquote{Active capture and
  stabilization of temporal solitons in microresonators,}
  {\protect\JournalTitle{Opt. Lett.}} \textbf{41}, 2037--2040 (2016).

\bibitem{Grudinin:11}
I.~Grudinin, H.~Lee, T.~Chen, and K.~Vahala, \enquote{Compensation of thermal
  nonlinearity effect in optical resonators,} {\protect\JournalTitle{Opt.
  Express}} \textbf{19}, 7365--7372 (2011).

\bibitem{Zhang:19}
S.~Zhang, J.~M. Silver, L.~D. Bino, F.~Copie, M.~T.~M. Woodley, G.~N. Ghalanos,
  A.~{\O}. Svela, N.~Moroney, and P.~Del'Haye, \enquote{Sub-milliwatt-level
  microresonator solitons with extended access range using an auxiliary laser,}
  {\protect\JournalTitle{Optica}} \textbf{6}, 206--212 (2019).

\bibitem{Zhou2019}
H.~Zhou, Y.~Geng, W.~Cui, S.-W. Huang, Q.~Zhou, K.~Qiu, and C.~Wei~Wong,
  \enquote{Soliton bursts and deterministic dissipative {K}err soliton
  generation in auxiliary-assisted microcavities,}
  {\protect\JournalTitle{Light: Science {\&} Applications}} \textbf{8}, 50
  (2019).

\bibitem{PhysRevApplied.12.034057}
G.~Moille, X.~Lu, A.~Rao, Q.~Li, D.~A. Westly, L.~Ranzani, S.~B. Papp,
  M.~Soltani, and K.~Srinivasan, \enquote{Kerr-microresonator soliton frequency
  combs at cryogenic temperatures,} {\protect\JournalTitle{Phys. Rev. Applied}}
  \textbf{12}, 034057 (2019).

\bibitem{Tanabe2016}
T.~{Kobatake}, T.~{Kato}, H.~{Itobe}, Y.~{Nakagawa}, and T.~{Tanabe},
  \enquote{Thermal effects on {K}err comb generation in a {CaF2}
  whispering-gallery mode microcavity,} {\protect\JournalTitle{IEEE Photonics
  Journal}} \textbf{8}, 1--9 (2016).

\bibitem{Guo2017}
H.~Guo, M.~Karpov, E.~Lucas, A.~Kordts, M.~H.~P. Pfeiffer, V.~Brasch,
  G.~Lihachev, V.~E. Lobanov, M.~L. Gorodetsky, and T.~J. Kippenberg,
  \enquote{Universal dynamics and deterministic switching of dissipative {K}err
  solitons in optical microresonators,} {\protect\JournalTitle{Nature Physics}}
  \textbf{13}, 94--102 (2017).

\bibitem{Joshi:16}
C.~Joshi, J.~K. Jang, K.~Luke, X.~Ji, S.~A. Miller, A.~Klenner, Y.~Okawachi,
  M.~Lipson, and A.~L. Gaeta, \enquote{Thermally controlled comb generation and
  soliton modelocking in microresonators,} {\protect\JournalTitle{Opt. Lett.}}
  \textbf{41}, 2565--2568 (2016).

\bibitem{Xue:16}
X.~Xue, Y.~Xuan, C.~Wang, P.-H. Wang, Y.~Liu, B.~Niu, D.~E. Leaird, M.~Qi, and
  A.~M. Weiner, \enquote{Thermal tuning of {K}err frequency combs in silicon
  nitride microring resonators,} {\protect\JournalTitle{Opt. Express}}
  \textbf{24}, 687--698 (2016).

\bibitem{Lobanov2015}
V.~E. Lobanov, G.~Lihachev, T.~J. Kippenberg, and M.~L. Gorodetsky,
  \enquote{Frequency combs and platicons in optical microresonators with normal
  {GVD},} {\protect\JournalTitle{Opt. Express}} \textbf{23}, 7713--7721 (2015).

\bibitem{PhysRevA.89.063814}
C.~Godey, I.~V. Balakireva, A.~Coillet, and Y.~K. Chembo, \enquote{Stability
  analysis of the spatiotemporal lugiato-lefever model for {K}err optical
  frequency combs in the anomalous and normal dispersion regimes,}
  {\protect\JournalTitle{Phys. Rev. A}} \textbf{89}, 063814 (2014).

\bibitem{Xue2015}
X.~Xue, Y.~Xuan, P.-H. Wang, Y.~Liu, D.~E. Leaird, M.~Qi, and A.~M. Weiner,
  \enquote{Normal-dispersion microcombs enabled by controllable mode
  interactions,} {\protect\JournalTitle{Laser \& Photonics Reviews}}
  \textbf{9}, L23--L28 (2015).

\bibitem{Jang:16}
J.~K. Jang, Y.~Okawachi, M.~Yu, K.~Luke, X.~Ji, M.~Lipson, and A.~L. Gaeta,
  \enquote{Dynamics of mode-coupling-induced microresonator frequency combs in
  normal dispersion,} {\protect\JournalTitle{Opt. Express}} \textbf{24},
  28794--28803 (2016).

\bibitem{Kim:19}
B.~Y. Kim, Y.~Okawachi, J.~K. Jang, M.~Yu, X.~Ji, Y.~Zhao, C.~Joshi, M.~Lipson,
  and A.~L. Gaeta, \enquote{Turn-key, high-efficiency {K}err comb source,}
  {\protect\JournalTitle{Opt. Lett.}} \textbf{44}, 4475--4478 (2019).

\bibitem{Lobanov2015epl}
V.~E. Lobanov, G.~Lihachev, and M.~L. Gorodetsky, \enquote{Generation of
  platicons and frequency combs in optical microresonators with normal {GVD} by
  modulated pump,} {\protect\JournalTitle{{EPL (Europhysics Letters)}}}
  \textbf{112}, 54008 (2015).

\bibitem{Lobanov2019}
V.~E. Lobanov, N.~M. Kondratiev, A.~E. Shitikov, R.~R. Galiev, and I.~A.
  Bilenko, \enquote{Generation and dynamics of solitonic pulses due to pump
  amplitude modulation at normal group-velocity dispersion,}
  {\protect\JournalTitle{Phys. Rev. A}} \textbf{100}, 013807 (2019).

\bibitem{KondratievNum:20}
N.~M. Kondratiev, V.~E. Lobanov, E.~A. Lonshakov, N.~Y. Dmitriev, A.~S.
  Voloshin, and I.~A. Bilenko, \enquote{Numerical study of solitonic pulse
  generation in the self-injection locking regime at normal and anomalous group
  velocity dispersion,} {\protect\JournalTitle{Opt. Express}} \textbf{28},
  38892--38906 (2020).

\bibitem{jin2020hertzlinewidth}
W.~Jin, Q.-F. Yang, L.~Chang, B.~Shen, H.~Wang, M.~A. Leal, L.~Wu, M.~Gao,
  A.~Feshali, M.~Paniccia, K.~J. Vahala, and J.~E. Bowers,
  \enquote{Hertz-linewidth semiconductor lasers using {CMOS}-ready
  ultra-high-{Q} microresonators,} {\protect\JournalTitle{Nature Photonics}}
  (2021).

\bibitem{Kondratiev:20}
N.~M. Kondratiev and V.~E. Lobanov, \enquote{Modulational instability and
  frequency combs in whispering-gallery-mode microresonators with
  backscattering,} {\protect\JournalTitle{Phys. Rev. A}} \textbf{101}, 013816
  (2020).

\bibitem{Savchenkov_2018}
A.~Savchenkov and A.~Matsko, \enquote{Calcium fluoride whispering gallery mode
  optical resonator with reduced thermal sensitivity,}
  {\protect\JournalTitle{Journal of Optics}} \textbf{20}, 035801 (2018).

\bibitem{Lim2017}
J.~Lim, A.~A. Savchenkov, E.~Dale, W.~Liang, D.~Eliyahu, V.~Ilchenko, A.~B.
  Matsko, L.~Maleki, and C.~W. Wong, \enquote{Chasing the thermodynamical noise
  limit in whispering-gallery-mode resonators for ultrastable laser frequency
  stabilization,} {\protect\JournalTitle{Nature Commun.}} \textbf{8}, 8 (2017).

\bibitem{FujiiTanabe2020}
S.~Fujii and T.~Tanabe, \enquote{Dispersion engineering and measurement of
  whispering gallery mode microresonator for {K}err frequency comb generation,}
  {\protect\JournalTitle{Nanophotonics}} \textbf{9}, 1087--1104 (2020).

\bibitem{Savchenkov:07}
A.~A. Savchenkov, A.~B. Matsko, V.~S. Ilchenko, and L.~Maleki, \enquote{Optical
  resonators with ten million finesse,} {\protect\JournalTitle{Opt. Express}}
  \textbf{15}, 6768--6773 (2007).

\end{thebibliography}

% Full bibliography added automatically for Optics Letters submissions; the following line will simply be ignored if submitting to other journals.
% Note that this extra page will not count against page length
\bibliographyfullrefs{sample}

\end{document}